\documentclass[sigconf]{acmart}

\AtBeginDocument{%
  \providecommand\BibTeX{{%
    \normalfont B\kern-0.5em{\scshape i\kern-0.25em b}\kern-0.8em\TeX}}}

\copyrightyear{2022}
\acmYear{2022}
\setcopyright{acmlicensed}\acmConference[MSR '22]{19th International Conference on Mining Software Repositories}{May 23--24, 2022}{Pittsburgh, PA, USA}
\acmBooktitle{19th International Conference on Mining Software Repositories (MSR '22), May 23--24, 2022, Pittsburgh, PA, USA}
\acmPrice{15.00}
\acmDOI{10.1145/3524842.3528501}
\acmISBN{978-1-4503-9303-4/22/05}

\usepackage{amsmath} 
\usepackage{listings}   
\usepackage{authorcomments}
\usepackage[ruled,vlined,linesnumbered]{algorithm2e}
\usepackage{algorithmicx}
\usepackage{algpseudocode}
\usepackage{subcaption}
\usepackage[export]{adjustbox}

\usepackage{hyperref}

\definecolor{javared}{rgb}{0.6,0,0} 
\definecolor{javagreen}{rgb}{0.25,0.5,0.35} 
\definecolor{javapurple}{rgb}{0.5,0,0.35} 
\definecolor{javadocblue}{rgb}{0.25,0.35,0.75} 

\lstdefinelanguage{JavaScriptColor}{
  keywords={await, async, break, case, catch, const, continue, debugger, default, delete, do, else,
    export, finally, for, function, if, import, in, instanceof, let, of, new, null, return, require, switch, this,
    throw, try, typeof, var, void, while, with, super,
    class, interface, implements, public, private, constructor
    },
  morecomment=[s]{/*}{*/},
  morestring=[b]',
  morestring=[b]",
  keywordstyle=\color{javapurple}\bfseries,
  identifierstyle=\color{black},
  commentstyle=\color{javagreen}\ttfamily,
  numberstyle=\color{javared}\ttfamily,
  stringstyle=\color{javared}\ttfamily,
  sensitive=false,
  numbers=left,
  stepnumber=1,
  escapeinside={/*\#}{\#*/},
}
\algnewcommand\algorithmicInput{\textbf{Input:}}
\algnewcommand\Input{\item[\algorithmicInput]}

\algnewcommand\algorithmicOutput{\textbf{Output:}}
\algnewcommand\Output{\item[\algorithmicOutput]}

\algnewcommand\algorithmicResult{\textbf{Result:}}
\algnewcommand\Result{\item[\algorithmicResult]}

\algnewcommand\algorithmicforeach{\textbf{for each}}
\algdef{S}[FOR]{ForEach}[1]{\algorithmicforeach\ #1\ \algorithmicdo}
\algdef{SE}[DOWHILE]{Do}{doWhile}{\algorithmicdo}[1]{\algorithmicwhile\ #1}

\algnewcommand\algorithmicpredicate{\textbf{predicate}}
\algdef{SE}[PREDICATE]{Predicate}{EndPredicate}%
   [2]{\algorithmicpredicate\ \textproc{#1}\ifthenelse{\equal{#2}{}}{}{(#2)}}%
   {\algorithmicend\ \algorithmicpredicate}%

\algnewcommand\algorithmicretpred{} 
\algdef{SE}[RETPRED]{RetPred}{EndRetPred}%
   [2]{\algorithmicretpred\ \textproc{#1}\ifthenelse{\equal{#2}{}}{}{(#2)}}%
   {\algorithmicend\ \algorithmicretpred}%

\newcommand{\tool}{\mbox{\textit{npm-filter}}\xspace}

\begin{document}

\title{\tool: Automating the mining of dynamic
information\\ from {\tt npm} packages}

\author{Ellen Arteca and Alexi Turcotte}
\authornote{Both authors contributed equally to this research.}
\email{{arteca.e, turcotte.al}@northeastern.edu}


\begin{abstract}
  The static properties of code repositories, e.g., lines of code, dependents, dependencies, etc. can be readily scraped from code hosting platforms such as GitHub, and from package management systems such as npm for JavaScript;
Although no less important, information related to the \textit{dynamic} properties of programs, e.g., number of tests in a test suite that pass or fail, is less readily available.
The ability to easily collect this dynamic information could be immensely useful to researchers conducting corpus analyses, as they could differentiate projects based on properties that can only be observed by running them.

In this paper, we present \tool, an automated tool that can download, install, build, test, and run custom user scripts over the source code of JavaScript projects available on npm, the most popular JavaScript package manager.
We outline this tool, describe its implementation, and show that \tool has already been useful in developing evaluation suites for multiple JavaScript tools.



\end{abstract}




\maketitle


%
%
%
%
%
%
\section{Introduction}

Many code hosting platforms contain a wealth of useful metadata: e.g., GitHub lists code authors, commits, and general project history, and library repositories (such as npm for JavaScript) often contain information on dependencies and dependents. 
Although it can be readily scraped from the web, this metadata is \emph{static}, and does not tell you much about running the actual code.
We thus define {\it dynamic} metadata to be information gleaned from program executions: e.g., number of running tests, code coverage of tests, performance, memory usage, etc.
Making said dynamic metadata available can enable new corpus analyses, focused on data pertaining to program executions---this is the purpose of our tool, \tool.

\tool is a tool for automatically installing, building, and testing sets of npm packages. 
npm is a major repository for JavaScript library code, and already contains a wealth of static metadata about JavaScript projects (and, if available, a link to the code).
\tool complements this by generating information such as how a project is built, if and how it is tested, the number of passing/failing tests, and the list of transitive dependencies.
\tool runs package code in a sandbox for added security and to ensure reproducibility of results.
Users can also specify custom scripts to run over the source code of the package.
As far as we know, there is no similar framework to automatically build, run, and test npm packages.

\tool has already been used to great effect in three projects; it was used to filter through huge lists of JavaScript projects in crafting evaluations for the \emph{DrAsync} anti-pattern detection tool~\cite{drasync}, the \emph{Nessie} test generator for asynchronous JavaScript callbacks~\cite{asyncLT}, and the \emph{Desynchronizer} tool for automatically migrating from synchronous JavaScript APIs to their asynchronous equivalents~\cite{desynchronizer}.

\section{Background \& Motivation}

Node.js~\cite{nodejs} is an eminently popular JavaScript runtime, particularly for server-side JavaScript, and while JavaScript is best known as a front-end, client-side language, it is rapidly gaining in popularity for server-side development~\cite{andreasen2017survey}.
npm~\cite{npm} is the most popular package ecosystem for Node.js applications: with the npm command-line interface (CLI) installed, a developer needs only navigate to the root of their project and {\tt npm install <package-name>} to download and install any package they desire.
JavaScript packages have a {\tt package.json} file in which users can specify commands that can be run by npm, e.g., many developers will specify a {\tt test} command that describes how a package's test suite is run, then a user can execute the package's tests with {\tt npm run test}.

npm provides a wealth of metadata for all of the projects it hosts, including the number of weekly downloads, dependencies, dependents, and a link to the associated code repository.
This said, running application code can reveal yet more useful information, such as if the package is equipped with a test suite, passing, failing, flaky tests, etc.
But even though it is relatively straightforward to install, build, and test an npm package, in our anecdotal experience conducting JavaScript tool evaluations, we found that only (roughly) <5\% of npm packages have running test suites with no failing tests. 

\tool can be used in any scenario where metadata about the execution of JavaScript code is required.
A list of npm projects or JavaScript repositories (e.g., from GHTorrent~\cite{Gousi13}, CodeDJ~\cite{maj_et_al:LIPIcs.ECOOP.2021.6}, or scraping npm), can be fed into \tool to gather dynamic metadata
by trying to install, build, and run package tests.

\section{\tool Design}



We will describe the overall design of \tool by way of describing the steps involved in analyzing a given npm package.
Analysis comprises various phases of execution, which correspond to the tasks required to set up and test an npm package, and running any user-specified scripts over the package's source code.

\subsection{Package Setup and Installation}

Supplied with an npm package name, \tool scrapes the repository link from the npm package page.
The source code is then downloaded (with a {\tt git clone}).
If the user specified a particular commit to be analyzed, then the source code is checked out at this commit.
If there is no repository link found on the page or if there is an issue with the cloning, then \tool bails out at this stage and reports the error to the user.

Once the code has been downloaded, the package dependencies are installed~\footnote{
\tool supports both {\tt npm} and {\tt yarn} package managers for installing dependencies.}.
The list of transitive dependencies can be a useful piece of data: for example, \cite{zimmermann2019small} show that transitive dependencies can contain vulnerabilities that compromise the package itself.
\tool computes this list by reporting the list of all packages in the {\tt node\_modules} directory after the install phase has completed.
There is also an option to exclude {\tt devDependencies}, which are dependencies excluded from production distributions of the package.




\subsection{Building a Package}

Once installed, some npm packages have additional commands that need to be run before the package is operational: we call this the \emph{build phase}.
For instance, packages written in TypeScript need to be compiled to JavaScript, and another common build step is the application of a bundler such as rollup~\cite{rollupNPM} or webpack~\cite{webpackNPM}.

To determine the build commands, \tool looks at the package's {\tt package.json} file and finds the available commands matching our tracked build commands.
By default, these are {\tt ``build''}, {\tt ``compile''}, and {\tt ``init''} (the most common build commands in our experience).
However, users can also customize the build commands tracked with a custom configuration file (discussed in Section \ref{configFileSection}).

If there is an error running a particular build command, the problematic command is added to the end of the command list; this way, the command can be run after potentially prerequisite commands. 
If all the build commands in a list have errors, then \tool bails out (to avoid infinite cycling) but continues to the testing phase anyway, reporting the build error in the results.


\subsection{Testing a Package}

Next, \tool determines if the package has a test suite, and if so computes some dynamic metadata---this is the \emph{test phase}.
{\tt package.json} is further parsed, this time to find the test commands.
By default, these are the common ones we observed: {\tt ``test'', ``unit' ', ``cov'', ``ci'', ``integration'', ``lint'', ``travis'', ``e2e'', ``bench'', ``mocha'', ``jest'', ``ava'', ``tap'', ``jasmine''}\footnote{
Many of these correspond to JavaScript testing infrastructures, such as {\tt mocha}.
}.

For each test command, \tool runs it and determines, by parsing the command itself and its output:
\begin{itemize}
    \item  if it is a linter or a coverage tool, and if so what tool is used;
    \item if not for linter/coverage, what testing infrastructure is used;
    \item whether or not it runs new user tests (this is false in test commands that only call other test commands, or that don't run any tests explicitly, e.g., linters, coverage tools);
    \item if it runs other test commands, then a list of these commands; 
    \item if it does run new user tests, then the number of passing and number of failing tests.
\end{itemize}

\tool parses the output of running tests with the following tools, that were the most common we observed in practice: eslint~\cite{eslintNPM}, tslint~\cite{tslintNPM}, xx~\cite{xxNPM}, standard~\cite{standardNPM}, prettier~\cite{prettierNPM}, gulp lint~\cite{gulpLintNPM} (linters); istanbul/nyc~\cite{istanbulNPM}, coveralls~\cite{coverallsNPM}, c8~\cite{c8NPM} (coverage tools); mocha~\cite{mochaNPM}, jest~\cite{jestDocs}, jasmine~\cite{jasmineNPM}, tap~\cite{tapNPM}, lab~\cite{labNPM}, ava~\cite{avaNPM}, gulp~\cite{gulpNPM} (test tools).
Any test commands that run other infrastructures (such as custom Node.js scripts) will still be parsed on a best-effort basis, and whether or not the correct number of passing/failing tests is determined depends on the shape of the output.

\subsection{Running Custom Scripts and CodeQL}

In addition to the metadata collected about the package build and test suite, users can also specify shell scripts and CodeQL~\cite{qlrepo} static analysis queries to be run over the source code of the package.
The scripts are run in the sequence specified, and any terminal output of each of them is included in the results, including errors.

CodeQL is a semantic code analysis language: with it, users can write static analyses for a variety of languages, including (most relevantly for \tool) JavaScript/TypeScript.
In Section~\ref{sec:ALTUsage}, we describe how this features was already used in an existing tool.


\subsection{Results}

The results of all phases of \tool are output to a JSON file.
This JSON results object is organized in a hierarchical structure corresponding to the aforedescribed phases of execution.
Any errors in an execution phase are reported in the corresponding field of the results.
The output file is named {\tt [package name]\_\_results.json}.

If the user specifies CodeQL queries to be run over the package source code, the output of each of these queries is output to a CSV file, named {\tt [package name]\_\_[query name]\_\_results.csv}.
Any errors in the CodeQL query execution would be reported in the CodeQL field of the JSON results.

\section{Implementation}
\tool is written in Python.
All the npm commands we run are done by dispatching with the Python {\tt subprocess} library; this allows us to parse the output, and specify a timeout.
It also doesn't crash \tool if there is any error in the subprocess.

The back end of \tool's npm package analyzer is a web scraper: given the name of an npm package, it finds the associated repository link on the npm page so that it can analyze the package's source code. 
The scraper is built using Python's scrapy library~\cite{scrapy}, which allows us to include custom middleware to run if the scraper gets an error code as a response from the site.
We implemented some middleware to deal with errors caused by the rate limiting on the npm site: if the site returns an error indicating that too many requests were received, the scraper pauses and then retries.
This middleware ensures that the scraper will not miss package information because of the rate limiter, but if a user is analyzing a large number of packages they will see a significant performance hit compared to running on the GitHub repos directly.
Thus, we also provide an option for users to pass a list of GitHub repos instead of npm packages to be analyzed, skipping the scraping entirely.

\tool is open source and includes a detailed Readme, with more examples than are included in this paper. 
\tool is available at \href{https://github.com/emarteca/npm-filter/}{\textcolor{blue}{https://github.com/emarteca/npm-filter/}}~\cite{zenodoLink}.


\section{\tool usage}
In this section, we explain how to use \tool and give some examples of usage.
To follow along, clone the source code linked above; all example commands are run from the root of the repo.
We have also included a minimal example usage tutorial \href{https://github.com/emarteca/npm-filter/blob/master/Tutorial.md}{\textcolor{blue}{here}}\footnote{
\href{https://github.com/emarteca/npm-filter/blob/master/Tutorial.md}{\textcolor{blue}{https://github.com/emarteca/npm-filter/blob/master/Tutorial.md}}
}.

\subsection{Safety first: Running in Docker}

\tool can be run in a docker container that is provided \href{https://hub.docker.com/r/emarteca/npm-filter}{\textcolor{blue}{on DockerHub}}\footnote{
\href{https://hub.docker.com/r/emarteca/npm-filter}{\textcolor{blue}{https://hub.docker.com/r/emarteca/npm-filter}}
}, and we recommend this usage.
The repository's Readme includes a list of all system requirements if you choose to run it locally or if you want to rebuild the docker container.
To run \tool sandboxed, simply preface any commands with {\tt ./runDocker.sh}.

\subsubsection{Input/Output from docker to host machine}
Running \tool in docker allows all the code being analyzed to be run in a sandbox, protecting the host machine. 
To allow input to \tool and access to the results files from running in docker, we have some special directories that the docker container has access to.
All input files to running \tool in docker must be in a directory {\tt docker\_configs} in the \tool home directory (any user scripts, CodeQL queries, or custom configuration files).
Results files end up in the {\tt npm\_filter\_docker\_results} directory, which is also in the \tool home directory.

\subsection{Basic usage}
This tool can either take JavaScript packages specified as GitHub repository links, or as npm packages.

To run \tool over GitHub repo links, use the following:

\begin{lstlisting}
./runDocker.sh python3 src/diagnose_github_repo.py 
		[--repo_list_file [rlistfile]] 
		[--repo_link [rlink]] 
		[--repo_link_and_SHA [rlink_and_SHA]] 
		[--config [config_file]]
                [--output_dir [output_dir]]
\end{lstlisting}

All arguments are optional, although \tool will not do anything if no repo links are specified. 
\begin{itemize}
    \item {\tt repo\_list\_file}: a file containing a list of GitHub repo links to be analyzed.
    Each line of the input file must specify one repo link, with an optional whitespace delimited commit SHA to check the repo out at. 
    \item {\tt repo\_link}: a link to a single GitHub repo to be analyzed
    \item {\tt repo\_link\_and\_SHA}: link to a GitHub repo followed by a space-delimited commit SHA to analyze the repo at 
    \item {\tt config}: path to a configuration file for the tool (config options explained in Section \ref{configFileSection})
    \item {\tt output\_dir}: path to a directory in which to output the results files (note: this only works when not running in docker)
\end{itemize}

To run \tool over npm packages, use the following:
\begin{lstlisting}
./runDocker.sh python3 src/diagnose_npm_package.py
			--packages [list_of_packages]
			[--config [config_file]]
			[--html [html_file]]
			[--output_dir [output_dir]]
\end{lstlisting}

\begin{itemize}
    \item {\tt packages}: list of npm packages to analyze. 
    Required argument, and at least one package must be passed.
    \item {\tt config}: path to a configuration file for the tool
    \item {\tt html}: path to an html file that represents the npm page for the package that is specified to be analyzed. This option only works for one package, so if you want to use this option on multiple packages you'll need to call the tool in sequence.
    \item {\tt output\_dir}: path to a directory in which to output the results files (note: this only works when not running in docker)
\end{itemize}

\subsubsection{Example Usage}

What follows is an example of basic usage.
This example runs on a single package, specified by GitHub repo and at a specific commit (to ensure consistency of expected output).
\begin{lstlisting}
./runDocker.sh python3 src/diagnose_github_repo.py
  --repo_link_and_SHA https://github.com/streamich/memfs
        863f373185837141504c05ed19f7a253232e0905
\end{lstlisting}

The results file is {\tt npm\_filter\_docker\_results/}\\ {\tt memfs\_\_results.json}, with contents (slightly redacted for length):
\begin{lstlisting}
    "installation": {
        "installer_command": "yarn"
    },
    "build": {
        "build_script_list": [
            "build"
        ]
    },
    "testing": {
        "test": {
            "num_passing": 265,
            "num_failing": 0,
            "test_infras": [
                "jest"
            ],
------------------------------ REDACTED FOR LENGTH
    "metadata": {
      "repo_link": "https://github.com/streamich/memfs",
      "repo_commit_SHA": REDACTED FOR LENGTH
    }
\end{lstlisting}
From this we can see that at this commit {\tt memfs} has a test suite with 265 passing tests and no failing tests, among other metadata.

More examples are included in the \tool GitHub repo Readme.



        

\subsubsection{Batch dispatch}
A common application of \tool is to analyze a large number of packages/repos.
We provide a bash script that dispatches \tool in parallel across batches of inputs.
\begin{lstlisting}
./runParallelGitReposDocker.sh repo_link_file
\end{lstlisting}
Results are in {\tt npm\_filter\_parallel\_docker\_results}. 
Note that this parallel execution in performed in one docker container, and not multiple parallel docker containers.

\subsection{Custom \tool configuration}\label{configFileSection}
Users can customize the behaviour of the tool by providing a custom configuration JSON file, organized by phases of \tool analysis. 
All fields are optional -- if not provided, defaults will be used\footnote{
Default configuration:  \href{https://github.com/emarteca/npm-filter/tree/master/configs}{\textcolor{blue}{https://github.com/emarteca/npm-filter/tree/master/configs}}.
}.

\subsubsection*{Install} package installation. 
\begin{itemize}
    \item {\tt timeout}: number of millisections after which, if the install is not complete, the process bails with a timed out error
\end{itemize}

\subsubsection*{Dependencies} package dependency tracking (this is the libraries the current package depends on, both directly and transitively). 
\begin{itemize}
    \item {\tt track\_deps}: specifies to compute the package dependencies
    \item {\tt include\_dev\_deps}: if true, this specifies to include the devDependencies in the dependency computation
    \item {\tt timeout}: timeout in milliseconds 
\end{itemize}

\subsubsection*{Build} package compile/build stage. 
\begin{itemize}
    \item {\tt tracked\_build\_commands}: any npm script with one of these listed commands as a substring will be tested. 
    \item {\tt timeout}: timeout in milliseconds, per build command
\end{itemize}

\subsubsection*{Test} package test stage. 
\begin{itemize}
    \item {\tt track\_tests}: specifies to run this testing diagnostic stage
    \item {\tt tracked\_test\_commands}: any npm script with one of these listed commands as a substring will be tested. 
    \item {\tt timeout}: timeout in milliseconds, per test command
\end{itemize}

\subsubsection*{Meta-info} any analysis-level configurations. 
\begin{itemize}
    \item {\tt VERBOSE\_MODE}: if true, include full output of all commands 
    \item {\tt ignored\_commands}: commands to ignore: if these are present in the npm script name, then they are not run even if they otherwise fall into a category of commands to run.
    \item {\tt ignored\_substrings}: commands to ignore: if these strings are present in the command string itself, then these npm scripts are not run (same as {\tt ignored\_commands}, but for the command strings instead of the npm script names)
    \item {\tt rm\_after\_cloning}: delete the package source code after the tool is done analyzing it.
    Strongly recommended if running over a large batch of packages.
    \item {\tt scripts\_over\_code}: list of paths to script files to run over the package source code. 
    \item {\tt QL\_queries}: list of paths to QL query files to run over the package source code.
\end{itemize}



%
%
%
%
%
%
\section{\tool in practice}

Now we describe three research papers that have used \tool.

%
%
%
%
\subsection{DrAsync}

Turcotte et al. used \tool to collect projects to evaluate their tool to detect anti-patterns in asynchronous JavaScript programs~\cite{drasync}.
Their tool, called \emph{DrAsync}, can statically detect asynchronous anti-patterns, and they found that many of these anti-patterns could be manually refactored; in order to confirm that these refactorings preserved behaviour, the authors ran application tests before and after refactoring (to confirm that refactoring did not introduce any failing tests).
The tool also has a dynamic component that records promise lifetimes and displays them in a visualization.

Thus, the evaluation undertaken in the paper requires running test suites, and \tool was used to filter a list of ~40K JavaScript Github repositories with asynchronous JavaScript code to a much more manageable 450 projects that had running/passing tests.
This work is being presented concurrently at ICSE Technical Track.

\subsection{Nessie}\label{sec:ALTUsage}

Arteca et. al built a test generator for JavaScript APIs with callback arguments~\cite{asyncLT}.
In this project, they wrote a static analysis in CodeQL, to identify pairs of nested calls to functions that were part of the APIs the test generator was targeting.
Then they used the CodeQL plugin feature of \tool to run this analysis on 13.6K JavaScript projects on GitHub.
The results of this CodeQL query, amalgamated across all 13.6K projects, was used to inform the test generator of common pairs of nested API calls, to generate tests more representative of developers' use of the APIs.
They also used \tool to select projects to evaluate the test generator.
This work is being presented concurrently at ICSE Technical Track.

\subsection{Desynchronizer}

Gokhale et al. used \tool to collect projects to evaluate their tool for automatically migrating projects that use synchronous JavaScript APIs to use their asynchronous equivalents~\cite{desynchronizer}.
The tool, called \emph{Desynchronizer}, statically detects calls to synchronous JavaScript APIs that have asynchronous equivalents (e.g., calls to {\tt readFileSync}, rather than {\tt readFile})---then infers a call graph, and automatically refactors the code.
In the evaluation, authors automatically applied {\it every} refactoring, and ran test suites post refactoring to establish any behavioural differences.

Thus, runnable test suites with no failing tests were required in the evaluation, and \tool was used to filter a list of 50K JavaScript projects using synchronous APIs targeted by the tool down to a few hundred projects with passing test suites.
\section{\tool Limitations}

We currently only support packages hosted on GitHub: if there is no GitHub repo link available on the package page, then \tool will not work. 
In our use cases we have found this to be rare. 

If the package uses a testing tool that we have not implemented output parsing for, then it might not be properly tracked.
That said, we have covered the most popular JavaScript test ecosystems.
Also, if the package uses build/test commands that don't include the substrings we expect, then they won't be run.
Note, however, that users can customize their \tool configuration to add or remove as many tracked commands as they want.
\section{Conclusion}

npm and GitHub contain a wealth of metadata related to static JavaScript project properties, but augmenting this static information with dynamic properties such as the number of tests in a test suite that pass or fail is immensely useful to researchers conducting corpus analyses or testing program transformation tools.
In this paper, we presented \tool, an automated tool that can download, install, build, test, and run custom user scripts over the source code of JavaScript projects available on npm, the most popular JavaScript package manager.
In addition to describing the implementation and usage of \tool, we also show that it has already been useful in developing evaluation suites for three separate JavaScript tools.

\begin{acks}
Both authors were supported in part by National Science
Foundation grants CCF-1715153 and CCF-1907727, and by the
Natural Sciences and Engineering Research Council of Canada.
\end{acks}

\bibliographystyle{ACM-Reference-Format}
\bibliography{paper}


\end{document}